\documentclass{elsart}

\usepackage{amssymb}
\usepackage{amsmath}
\usepackage{amsfonts}
\usepackage[dvips]{graphicx}

\begin{document}

\begin{frontmatter}


% \address{}

\title{Wide bandwidth phase-locked diode laser with an intra-cavity electro-optic modulator}

%\author{J. Le Gou\"{e}t\corauthref{cor1}\thanksref{label1}, J. Kim\thanksref{label2}, C. Bourassin-Bouchet, M. Lours, A. Landragin, F. Pereira Dos Santos}
\author{J. Le Gou\"{e}t\corauthref{cor1}\thanksref{label1}},
\author{J. Kim\thanksref{label2}},
\author{C. Bourassin-Bouchet},
\author{M. Lours},
\author{A. Landragin},
\author{F. Pereira Dos Santos}

%\ead{le\_gouet@mit.edu}
\corauth[cor1]{le\_gouet@mit.edu}

\address{LNE-SYRTE, CNRS UMR 8630, UPMC,
Observatoire de Paris, 61 rue de l'Observatoire, 75014 Paris}
\address[label1]{Present address :
Research Laboratory of Electronics,
Massachusetts Institute of Technology,
77 Massachusetts avenue,
USA}
\address[label2]{Department of Physics,
Myongji University,
449-728 Yongin,
KOREA
}

\begin{abstract}
Two extended cavity laser diodes are phase-locked, thanks to an intra-cavity electro-optical modulator. The phase-locked loop bandwidth is on the order of 10~MHz, which is about twice larger than when the feedback correction is applied on the laser current. The phase noise reaches -120~dBrad$^2$/Hz at 10~kHz. This new scheme reduces the residual laser phase noise, which constitutes one of the dominant contributions in the sensitivity limit of atom interferometers using two-photon transitions.
\end{abstract}

\end{frontmatter}

% main text
\section{Introduction}
Potentialities of optical communications or opto-electronic techniques for detection has both risen interest for optically carried microwave signals. These technologies take benefit as well as they contribute to the development of optical spectroscopy and fundamental metrology. Laser stabilization \cite{Hall78} has for example allowed progress in time/frequency metrology \cite{Rosenband}, or led to the development of velocity and range measurements by optical sources, so called Lidar-Radar \cite{Eberhard,Mullen,Morvan}. Phase locked loop, in particular, has been long used in telecommunication with microwave electronics \cite{Blanchard}, before being successfully applied to realize low phase noise optical sources \cite{Steele}. We present here the application of a technique, originally used for optical communications \cite{Malyon}, to phase lock two independent laser diodes, which will allow coherent manipulation of atomic wave packets for atom interferometry.

Among the various coherent splitting processes for atom interferometry, two photons transitions are the most mature. Many
experiments based on stimulated Raman or Bragg transitions have been reported so far \cite{Kasevich91,Giltner,Rasel}, and they are the most common
way to realize high sensitivity measurements with atom interferometry \cite{Gustavson00,Peters01,Vigue07,Kasevich072,Tino08,Canuel,LeGouet08}. These
interferometers are based on the measurement of the atom displacement with respect to the equiphases defined by two light waves.
A stable phase relation is thus required between the two beams. For stimulated Raman transitions, this relation can be provided by
servo locking two independent laser sources, with a low noise phase-locked loop.

In our experiment, the phase difference between two lasers is servo locked by comparing their beat-note to the signal delivered by a microwave oscillator. These laser sources for the Raman transitions are two extended cavity diode lasers (ECDL). In general, the phase difference is stabilized by combinating a slow but broadband servo lock on the cavity length, and a fast correction obtained by controling the current of the diode \cite{Steele,Santarelli94}.

The residual noise on the phase difference between the two lasers is limited by the electronic bandwidth of the phase-locked loop (PLL). When applying corrections on the diode current, even though they are directly applied on the laser diode, the bandwidth barely exceeds 5 MHz, mainly limited by the residual capacitance of the pin connections of the
diode itself. Alternatively, fast corrections can be applied on the effective
cavity length, using an intra-cavity element whose refractive index can be electronically driven, such as an electro-optical modulator (EOM).
This scheme allows faster corrections (the intrinsic capacitance of the EOM is lower) and is independent of the diode chip, allowing to reach the same
bandwidth regardless to the wavelength and casing. Such an intra-cavity EOM has already been used to phase lock two ECDL at 1.5~$\mu$m \cite{Hyodo}, or two Titanium Sapphire lasers \cite{Muller06}, as well as to realize high bandwidth frequency locks \cite{Telle,Celikov,Crozatier}.

% Ajouter quelques ref sur freq lock avec EOM intracavité

In this paper, we report on the use of an intra-cavity EOM in a extended cavity diode laser in the near infrared range, in order to phase lock two such ECDL. This development is motivated by the need of reducing the impact of the laser phase noise on the phase stability of a cold atom gravimeter \cite{LeGouet08},
where alkali atom wavepackets are coherently split and recombined with stimulated Raman transitions. The PLL scheme that we present here reaches better performances than previously reported for laser diodes \cite{Hyodo, LeGouet08}. In particular, we improve considerably the bandwidth, up to about 10~MHz, and allow reducing the residual phase noise contribution to the interferometer sensitivity.

\section{Phase-locked loop setup}

Our atom interferometer is realized with $^{87}$Rb atoms. The optical setup is based on two linear ECDL,
emitting at a wavelength of 780~nm. The diodes are based on the design described in \cite{Baillard06}. Each cavity contains a laser diode chip (Sharp GH0781JA2C), a collimating lens, and a low-loss
interference filter as a frequency selective element. It is closed by a cat's eye, constituted by a converging lens and a 30\% reflection plate glued on a PZT actuator. The laser field emitted by the diode is linearly polarized, with a
polarization extinction ratio of 20/1 between the two orthogonal axes. The bare laser chip provides about 120~mW of optical power at
780~nm for a 150~mA pumping current. Because of the frequency selective filter and losses in the extended cavity, the output power of the ECDL is significantly less. Operating the laser at a moderate current of 90 mA, we get 30 mW, which is enough for our applications. The standard design corresponds to a 10~cm long linear cavity, which is used here for the reference laser.

The cavity of the phase-locked laser is larger: it contains the same above components and an additional 12~cm long phase modulator (Linos
PM25IR), which is inserted between the filter and the cat's eye. The extraordinary axis of the electro-optic modulator (EOM) is aligned with the
linear polarization of the laser diode, so that the effective optical length of the cavity can be modified by linear Pockels effect. The phase modulator
is constituted by two KDP crystals shaped at Brewster angle. The bandwidth of the phase modulator is rated to be as high as 100 MHz. Though we have found many small resonances, starting from acoustic frequencies up to more than 10 MHz, none of them could make the lock loop unstable. Two Brewster cut protection windows are mounted at the ends of the modulator. The absorption is lower than 1\% in the infrared range, so the threshold current is similar for both cavities (around 30 mA) and an optical power of 30~mW is preserved at the output of the phase-locked laser cavity. Despite an increased length of the cavity (20~cm long), the temperature and mechanical stability of the laser is comparable to the ECDL without intra-cavity EOM.

   \begin{figure}[h!]
        \centering
        \includegraphics[width=8 cm, angle=-90]{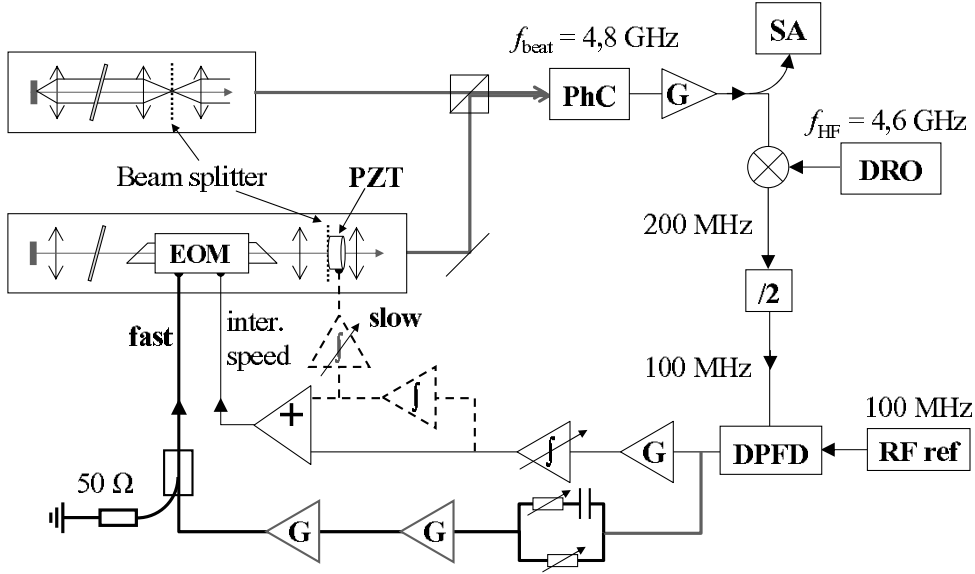}
\caption{Schemes of the linear ECDLs and electronic setup of their phase-lock loop. A slow loop reacts on the PZT actuator holding the
output coupler of one laser, while intermediate speed and fast corrections are respectively applied on the two electrodes of the
intra-cavity electro-optical modulator (EOM). SA: signal analyzer, DRO: Dielectric Resonator Oscillator, DPFD: Digital Phase Frequency Detector, EOM: electro-optic modulator, PZT: piezo electric transducer, PhC: photoConductor.}
        \label{setup}
   \end{figure}

The PLL experimental setup is sketched on figure \ref{setup}. The microwave (MW) beat-note of the two lasers is collected on a
photoconductor (PhC), and amplified. For this experimental demonstration, the amplified beat-note is mixed down with a Dielectric Resonator
Oscillator (DRO), at $f_\text{HF}=4.6$~GHz. In fact, the energy difference between the hyperfine ground states of the $^{87}$Rb atom corresponds to a frequency of 6.835~GHz, but the
noise level of the lasers phase difference does not depend on this frequency difference, so the present results should transpose readily at the $^{87}$Rb hyperfine splitting frequency. The downconverted signal at intermediate frequency $f_\text{IF}$ is then amplified. Its frequency is divided by a factor 2 and compared to an external 100~MHz reference (IFR~2023A) with a Digital Phase Frequency Detector (DPFD, model MCH 12140), whose output provides the error signal for the phase-locked loop.

As for the phase-locked loop electronics, we use a circuit similar to the one presently used in our atomic gravimeter, where the feedback is applied directly on the
laser diode current \cite{Cheinet_APB}. Here however, the voltage output of the gain electronics drives the EOM electrodes. The intermediate speed correction, obtained by integrating the error signal (see fig. \ref{setup}), is applied on one electrode of the EOM. The gain of this loop is increased at low frequencies by integrating once more the error signal and adding it to the correction signal. The second electrode is first grounded. A third integration provides the slow correction, which is applied on the piezo-electric actuator. This initial two loops circuit is thus identical to the one we usually used for the current supply feedback.

We then add a third correction to improve the bandwidth of the phaselock (thick lines in fig. \ref{setup}). A simple passive RC filter
provides the lead compensation of the DPFD signal, in order to partially compensate the phase shifts induced by the various amplifiers. Finally the signal is amplified by 30~dB
with two low-noise RF amplifiers (ZFL-500-LN and ZHL-3A) and applied to the second EOM electrode via a $50~\Omega$ adapted coupler, ensuring RF impedance matching.

\section{Residual noise of the phase-locked loop}
To study the influence of these various loops on the phase noise of the PLL, we measure the corresponding power spectrum densities (PSD).
For frequencies higher than 100~kHz, the PSD is directly measured on a fraction of the amplified beat-note, after the photoconductor, with
a spectrum analyzer. To analyze the phase noise at lower frequencies, the 100~MHz converted beat-note is mixed with the 100~MHz reference used in the PLL, and the spectrum of phase fluctuations is obtained with a FFT analyzer (not shown on the figure).
This measurement is free from the noise of the reference oscillator, hence revealing the residual noise of the PLL.

\begin{figure}[h!]
        \centering
        \includegraphics[width=10 cm, angle=0]{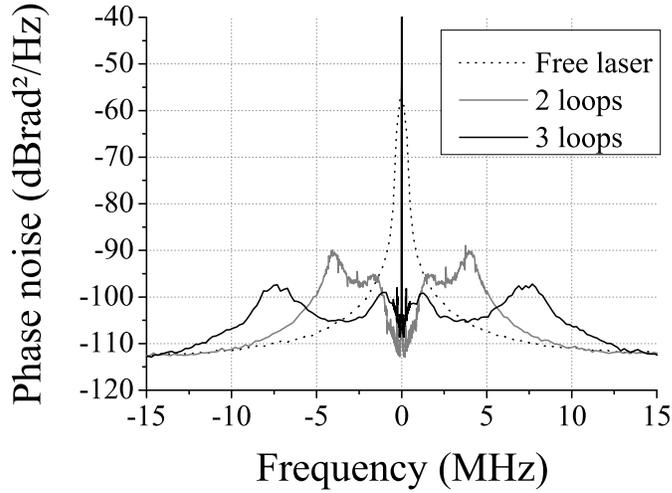}
\caption{Phase noise power spectral densities, measured on a spectrum analyser, of the optical beat-note between the two lasers, for the different possible configurations of the
phase-locked loop.}
        \label{PSD_lin}
   \end{figure}

The results from the spectrum analyzer are presented on figure \ref{PSD_lin}. The free running laser PSD (dotted line) is measured by closing only the slow loop on the PZT, whose gain is negligible for frequencies higher than 1 kHz. Adding the intermediate speed correction, we obtain a PLL bandwidth of up to 6~MHz. This is already larger than the bandwidth obtained when the feedback is applied onto the current supply (typically 1.5 and 4 MHz, respectively without and with lead compensation). This shows that the various delays due to the electronics (amplifiers and cables) are not the limitation of the PLL bandwidth with the diode current feedback. Closing the fast feedback loop, the bandwidth can reach between 8 and
12~MHz, depending on the integrators gains (fig. \ref{PSD_lin}).

\begin{figure}[h!]
        \centering
        \includegraphics[width=10 cm, angle=0]{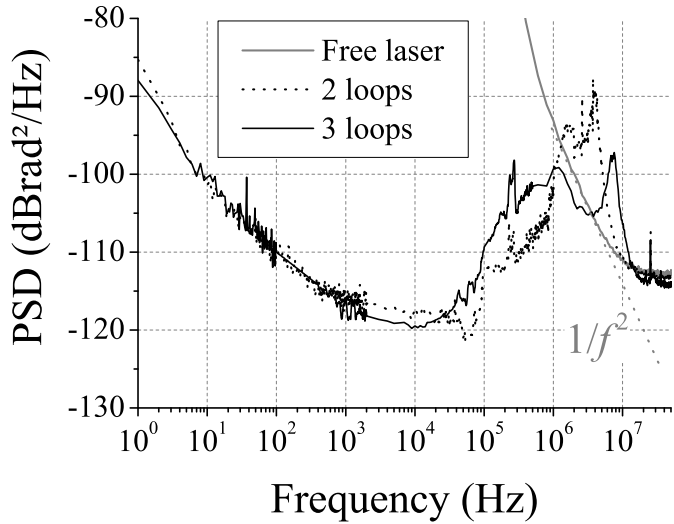}
\caption{Phase noise power spectral densities of the optical beat-note between the two lasers, when the integrator gains are optimized to
reduce the phase noise.}
        \label{PSD_log}
   \end{figure}

To reduce the phase noise within the bandwidth, the gain of the first
integrator in the intermediate speed loop is increased, which adds some phase shift at high frequency and reduces the bandwidth. This phase shift can be partly compensated for by adjusting the lead compensation circuit of the fast loop, looking for a compromise between gain and bandwidth. The black curve displayed on figure \ref{PSD_log} corresponds to such a compromise. Between 1~MHz and 6~MHz, the phase noise is reduced by up to 15 dB by the fast loop, compared to the situation with only slow
and intermediate speed loops. Between 300~kHz and 1~MHz, however, the fast loop configuration phase noise is up to 6~dB higher. The measurement of the phase noise of the lasers is limited by the intrinsic noise of the spectrum analyser for frequencies higher than 8~MHz. The phase noise reaches a plateau at -113~dBrad$^2$, whereas it should decrease much lower. Indeed, beyond the PLL bandwidth, the phase noise is not corrected anymore and corresponds to the free running laser, thus the real spectra for the three configurations should correspond to the caracteristic $1/f^2$ curve of a white phase noise (gray doted line).

Ultimately, the PLL bandwidth is limited by the electronic delays in the cables and the amplifiers in the loop. Compared to applying the correction on the diode laser current, the gain of the amplifiers of the electronic loops is much larger, because of the relatively small tuning sensitivity of the EOM feedback. It seems then that the main limitation arises from the limited gain-bandwidth product of the amplifiers. As for the residual noise at low frequency (below 10~kHz), it is limited by the intrinsic noise of the various electronic components. We reach a residual phase noise level close to $-120$~dBrad$^2$/Hz, from 1~kHz to 100~kHz.

We can compare further the performances of the PLL with intra-cavity EOM or laser current feedbacks by calculating their respective contribution to the noise of the interferometer. The influence of the laser phase noise can be quantified by weighting the power spectral density of the lasers phase difference fluctuations by a transfer function, derived from the formalism of the sensitivity function \cite{CheinetIEEE}. In particular, this transfer function acts as a low pass filter, with a cut off frequency on the order of the Rabi frequency of the Raman pulses. This formalism was used to evaluate the present sensitivity of our atomic gravimeter, which is presently limited by vibration noise to $1.4 \, 10^{-8}~\text{g}/\sqrt{\text{Hz}}$ \cite{LeGouet08}, for a duration of the interferometer of 100~ms, and a duration of the atomic beamsplitting pulses $\tau_{\pi/2}=8~\mu$s. The contribution of the residual phase noise of the PLL based on the laser diode current feedback is found to be $2 \, 10^{-9}~\text{g}/\sqrt{\text{Hz}}$. With the PLL based on the intra-cavity EOM feedback, and the typical spectra reported in the previous figures, the PLL contribution would be $1.5 \, 10^{-9}~\text{g}/\sqrt{\text{Hz}}$ either with the 2 or 3 loops setups, which represents a relatively small gain on the contribution. For shorter pulses, however, the effect of the EOM feedback would be more appreciable. Let us consider for example $\tau_{\pi/2}=1~\mu$s, meaning that the Rabi frequency must be $f_R=250$~kHz, and the cut-off frequency of the sensitivity function filtering $f_c=144$~kHz. In that case, the contribution of the residual phase noise of the PLL to the measurement sensitivity would be of $18 \, 10^{-9}~\text{g}/\sqrt{\text{Hz}}$ for the laser diode current feedback, only $9 \, 10^{-9}~\text{g}/\sqrt{\text{Hz}}$ for the EOM feedback with 2 loops, and $13 \, 10^{-9}~\text{g}/\sqrt{\text{Hz}}$ for the EOM feedback with 3 loops. The better figure of the 2 loops scheme is due to the fact that the laser phase noise of the 3 loops scheme is higher in the region of the cut-off frequency.

Lower phase noise can be obtained by using an analog mixer instead of a digital phase frequency detector, as shown in \cite{Muller06}. It may also be interesting to combine digital and analog phase and frequency detectors, to exploit their respective broad capture range and low noise performance \cite{Tino05}. However, the level of residual noise obtained here in the range 1-100~kHz is very close to the phase noise of the reference microwave signal which we have developed to be used as a phase reference at 6.8~GHz \cite{Cheinet_APB}. There is thus no point in drastically increasing the gain or reducing the intrinsic noise of the loop in this frequency range.

%Careful optimization of the different loops could help reducing the noise above 100 kHz. A enlever

As the power in the Raman beams in an atom interferometer is typically on the order of 100~mW, diode lasers have to be amplified. We verified that the performance of this phase-locked loop would be preserved when amplifying the diode lasers by injecting tapered amplifiers \cite{pipo_gyro}.

Finally, one could object that our measurements do not take into account the noise sources which are common to the two beams, like the photoconductor, or the vibrations in the beat setup. However, an independent measurement performed with two independent photoconductors showed that these contributions lie at -130 dBrad$^2$/Hz at 100 kHz, which is well below the residual noise obtained here, and are thus negligible. As for the various noise sources due to the propagation of the laser beams (index fluctuations in fibers, vibrations of the optics, propagation delay), their independent contributions have already been studied independently in previous works \cite{Cheinet_APB,LeGouet07,LeGouet08}.
%, and show that they were lower than the residual PLL contribution.

\section{Conclusion}
We reported here a low noise phase-lock of two independent extended cavity diode lasers, based on the control of the cavity
optical length. The intermediate speed and fast corrections are applied on an intra-cavity electro-optical modulator, so that the supply
current is not affected for the PLL. This technique allows here to reach a PLL bandwidth on the order of 10~MHz. To our knowledge, this bandwidth is the best ever
reported for a PLL between two diode lasers. Such a high bandwidth should also allow reaching an excellent frequency agility of the phase locked diode laser \cite{Muller06}, and reduce transient effects when closing the PLL loop \cite{Cheinet_APB}. Finally, our scheme allows reducing efficiently the relatively large phase noise of ECDL diode lasers.

A white noise floor of -120~dBrad$^2$/Hz at 10~kHz is obtained after optimizing the gain of the loop.
This setup will then be applied to drive the stimulated Raman transitions that realize our
cold atom gravimeter. A reduction of the residual PLL phase noise will reduce the corresponding contribution in the interferometer phase
noise, which is presently the higher one among the environment independent contributions \cite{LeGouet08}.

\section{Acknowledgement}
We would like to thank L. Volodimer for his help in the many modifications of the electronic setup, the Institut Francilien
pour la Recherche sur les Atomes Froids (IFRAF) and the European Union (FINAQS) for financial support. J.L.G. thanks the DGA for supporting
his work. J.K. thanks PHC-STAR for supporting his visit.

% \appendix

% \section{}
% \label{}
\bibliographystyle{unsrt}
%\addcontentsline{toc}{chapter}{Bibliographie}
\bibliography{EOM_Review}

\begin{thebibliography}{10}
\bibitem{pipo_gyro}
Measurements have been performed on the cold atom gyrometer of the
  \text{LNE-SYRTE}, to compare the phase noises before and after a \text{1~W}
  tapered power amplifier. \text{It} demonstrated that the contribution of the
  amplifier to the \text{PLL} residual phase noise lie below
  $-130$~\text{dBrad}$^2$/\text{Hz} above 10~\text{kHz}, and is thus
  negligible. \text{D}etails will be published in another article.
\end{thebibliography}

%\begin{thebibliography}{00}
%% notes:
%% \bibitem{label} \note
%
%% subbibitems:
%% \begin{subbibitems}{label}
%% \bibitem{label1}
%% \bibitem{label2}
%% If there is a note, it should come last:
%% \bibitem{label3} \note
%% \end{subbibitems}
%
%\bibitem{}
%
%\end{thebibliography}

\end{document}